\documentclass{pasj00}
\draft

\begin{document}
\SetRunningHead{Li \& Cao}{BL Lac Objects and the EGRB}

\title{BL Lacertae Objects and the Extragalactic $\gamma$-Ray
Background}

\author{
   Fan \textsc{Li}\altaffilmark{1,2}
   and
   Xinwu \textsc{Cao}\altaffilmark{1}}
 \altaffiltext{1}{Key Laboratory for Research in Galaxies and Cosmology,
Shanghai Astronomical Observatory, Chinese Academy of Sciences, 80
Nandan Road, Shanghai, 200030, China}
 \altaffiltext{2}{Graduate School of the Chinese Academy of Sciences,
Beijing 100039, China}
 \email{lifan@shao.ac.cn}

%


\maketitle

\begin{abstract}
A tight correlation between $\gamma $-ray and radio emission is
found for a sample of BL Lacertae (BL Lac) objects detected by {\it
Fermi Gamma-ray Space Telescope} (Fermi) and the Energetic Gamma-Ray
Experiment Telescope (EGRET). The $\gamma$-ray emission of BL Lac
objects exhibits strong variability, and the detection rate of
$\gamma$-ray BL Lac objects is low, which may be related to the
$\gamma$-ray duty cycle of BL Lac objects. We estimate the
$\gamma$-ray duty cycle, $\delta_{\gamma}\simeq 0.11$, for BL Lac
objects detected by EGRET and Fermi. Using the empirical relation of
$\gamma$-ray emission with radio emission and the estimated
$\gamma$-ray duty cycle $\delta_{\gamma}$, we derive the
$\gamma$-ray luminosity function (LF) of BL Lac objects from their
radio LF. Our derived $\gamma$-ray LF of BL Lac objects can almost
reproduce that calculated with the recently released Fermi bright
active galactic nuclei (AGN) sample. We find that  $\sim 45\%$ of
the extragalactic diffuse $\gamma$-ray background (EGRB) is
contributed by BL Lac objects. Combining the estimate of the quasar
contribution to the EGRB in the previous work, we find that $\sim
77\%$ of the EGRB is contributed by BL Lac objects and radio
quasars.
\end{abstract}

\section{Introduction}

There are 66 high-confidence identifications of blazars in the third
catalog of active galactic nuclei (AGN) detected by the Energetic
Gamma-Ray Experiment Telescope (EGRET) on the {\it Compton Gamma-Ray
Observatory (CGRO)}, which include 20 BL Lac objects (e.g.
\cite{har99,2001ApJS..135..155M}). The LAT bright AGN Sample (LBAS)
detected by Fermi includes 104 blazars consisting of 42 BL Lac
objects, 57 flat spectrum radio quasars (FSRQs), and 5 blazars with
uncertain classification \citep{a09a}. Comptonization
is believed to be responsible for the $\gamma$-ray emission from
blazars, which can be classified into two categories: the external
Comptonization (EC) model and the synchrotron self-Comptonization
(SSC) model, according to the origin of the soft seed photons (see,
e.g., \cite{2007ApSS.309...95B} for a review and references therein
). In the EC models, the soft seed photons are assumed to originate
from the accretion disks, the broad-line regions (BLRs), or/and the
dust tori (e.g.,
\cite{1996MNRAS.280...67G,2001ApJ...561..111G,2002ApJ...575..667D}).
It was suggested that the SSC model may be responsible for
$\gamma$-ray radiation from BL Lac objects while the EC model
predominates over SSC model for quasars (e.g., \cite{don95}). The
statistical analysis on a sample of EGRET blazars implied that the
soft seed photons may be predominantly from the BLRs (e.g.,
\cite{2004ApJ...602..103F,2006ApJ...646....8F}), which is consistent
with the lack of BLR and other external fields emission in most BL
Lac objects. Their results are roughly consistent with the detailed
modeling of SEDs for a large sample of $\gamma$-ray bright AGNs by
\citep{ghi10}, in which all external field photons from BLR,
accretion disk, and dust torus, are properly considered. However,
the situation becomes more complicated for radiation in other
wavebands (e.g., see
\cite{2005AIPC..745..522K,2009ApJ...699.2002B,2008ApJ...688..148L}).
It was found that the X-ray radiation from some FSRQs (e.g., 3C~273)
is consistent with combined SSC and EC mechanisms (e.g.,
\cite{p09}).

The extragalactic diffuse $\gamma$-ray background (EGRB) was first
discovered by the SAS 2 satellite \citep{fic78,tho82} and
subsequently confirmed after the launch of EGRET
\citep{mic95,sre98}. The EGRB integrated above 100 MeV was
determined to be $(1.45\pm 0.05)\times 10^{-5}$ photons cm$^{-2}$
s$^{-1}$ sr$^{-1}$ \citep{sre98}. Using a new model of the Galactic
background, \citet{str04} obtained a slightly smaller value of the
EGRB, $(1.14\pm 0.12)\times 10^{-5}$photons cm$^{-2}$ s$^{-1}$
sr$^{-1}$. Almost all extragalactic $\gamma$-ray sources in the
third EGRET catalog are identified as blazars, which account for
about 13\% of the EGRB \citep{har99}. The contribution of unresolved
blazars to the EGRB has been explored in many previous works, either
by extrapolating the observed $\gamma$-ray luminosity distribution
to obtain a $\gamma$-ray luminosity function (LF) or using the
correlation of $\gamma$-ray emission with the emission in radio
bands derived from EGRET blazars, which showed that about $\sim$25\%
to $\sim$100\% of the EGRB can be attributed to the unresolved
blazars (e.g., \cite{pad93,chi95,ste96,chi98,muc00,nar06,cao08}).



A new revised catalogue of EGRET $\gamma$-ray sources (EGR2008) was
given by \citet{cas08}, and a larger catalogue of the LAT Bright AGN
Sample (LBAS) detected by Fermi was released recently \citep{a09a}.
In this paper, we derive a correlation between $\gamma$-ray and
radio emission for a sample of BL Lac objects detected by EGRET
or/and Fermi \citep{cas08,a09a}. We estimate the $\gamma$-ray duty
cycle of BL Lac objects detected by EGRET and Fermi, and the
$\gamma$-ray LF of BL Lac objects is calculated with this empirical
correlation from their radio LF and $\gamma$-ray duty cycle. The
radio LF of BL Lac objects derived by \citet{pad07} based on the
Deep X-ray Radio Blazar Survey (DXRBS) is used in this work. We
further estimate the total number of BL Lac objects to be detected
by Fermi, and the contribution of all BL Lac objects to the EGRB.
The cosmological parameters $\Omega_{\rm M}=0.3$,
$\Omega_{\Lambda}=0.7$, and $H_0=70~ {\rm km~s^{-1}~Mpc^{-1}}$ have
been adopted in this paper.



\begin{figure}
  \begin{center}
    \FigureFile(80mm,80mm){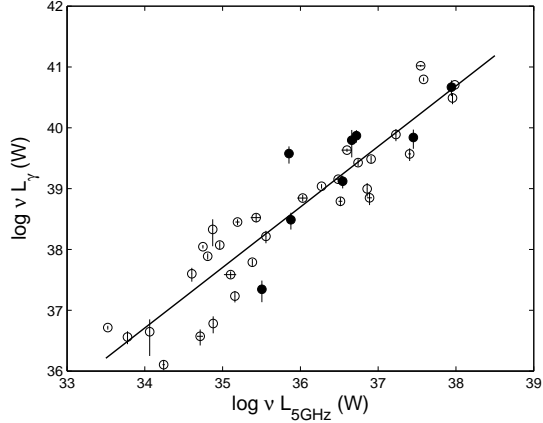}
  \end{center}
  \caption{The relation of $K$-corrected radio luminosity and
$\gamma$-ray luminosity for a sample of $\gamma$-ray BL Lac objects
detected by Fermi/EGRET. The solid line represents the linear
regression for all the sources in the sample. Open circles: Fermi BL
Lac objects; filled circles: EGRET BL Lac objects.}\label{fig1}
\end{figure}

\section{The correlation between $\gamma$-ray and radio emission}

The recently released LAT Bright AGN Sample (LBAS) detected by Fermi
includes 42 BL Lac objects, of which 11 sources were already
detected by EGRET. Till now, there are 51 BL Lac objects detected
either by EGRET or Fermi [with the Test Statistic (TS) value above
the EGRET threshold ($4\sigma$ for $\mid b\mid
>10\textordmasculine $, and 5$\sigma$ for $\mid b\mid
<10\textordmasculine $); or 10$\sigma$ for the sources detected by
Fermi]. There are 40 BL Lac objects with measured redshifts, which
are adopted for our present investigation. The radio/$\gamma$-ray
flux and photon index $\Gamma$ above 100MeV for the sources are
listed in Table 1.  In figure \ref{fig1}, we plot the relation
between radio luminosity $L_{\rm R,5G}$ at 5~GHz and $\gamma$-ray
luminosity $L_{\gamma}$ ($\ge 100~{\rm MeV}$) for these BL Lac
objects, where the average K-corrected radio and $\gamma$-ray fluxes
are adopted for calculating $L_{\gamma}$ and $L_{\rm R,5G}$.
We find a significant correlation between $L_{\rm R,5G}$ and
$L_{\gamma}$ (the correlation coefficient $r=0.926$). The linear
regression of the sample gives
\begin{equation}
\log \nu L_{\gamma}=0.995\log \nu L_{\rm R,5G}+2.886. \label{lr
lgam}
\end{equation}
\citet{a09a} investigated the relation between 8.4GHz radio flux
density and $\gamma$-ray flux density for Fermi blazar sample. The
peak values are adopted by \citet{a09a}, which are likely to
correspond to a short time-scale flare state of a source.
\citet{kov09} investigated the relation between quasi-simultaneous
$\gamma$-ray and 15GHz radio flux densities in Fermi blazar sample,
which includes xx BL Lac objects. The mean fluxes above 100~MeV are
adopted in this work, and the sample is limited to BL lac objects,
which should be a suitable choice, because we focus on the statistic
properties of $\gamma$-ray BL Lac objects as a whole population in
this work. The significance of the correlation found in this work is
higher than that found by \citet{a09a}, which may be attributed to
the fact that only the peak $\gamma$-ray fluxes were adopted in
their analysis. For $\gamma$-ray quasars, the situation is more
complicated, as their $\gamma$-ray emission is not only dependent on
their radio emission, but also on the origins of the soft seed
photons, which is beyond the scope of this work.

\begin{figure}
  \begin{center}
    \FigureFile(80mm,80mm){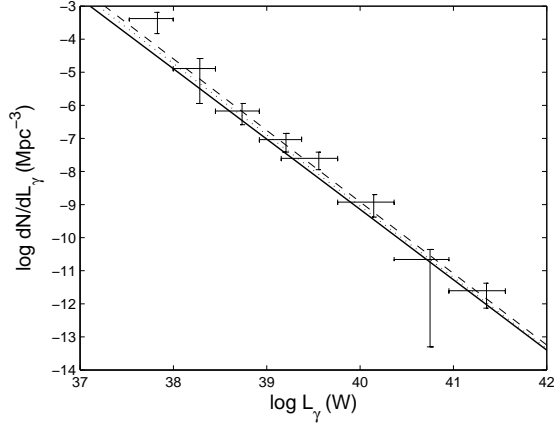}
  \end{center}
  \caption{The $\gamma$-ray LFs for BL Lac objects. The solid line is
the LF derived in this work, while the dashed line and the dotted
line are the LFs (redshift less than 0.3 and greater than 0.3)
calculated with the LAT bright AGN sample detected by Fermi
\citep{a09a}. }\label{fig2}
\end{figure}

\section{The $\gamma$-ray luminosity function of BL Lac objects}

Based on the correlation between $L_{\rm R,5G}$ and $L_{\gamma}$
established with a sample of BL Lac objects, we can calculate the
$\gamma$-ray LF of BL Lac objects from their radio LF. Recently,
\citet{pad07} derived a LF for BL Lac objects based on the Deep
X-ray Radio Blazar Survey (DXRBS), which extends to lower luminosity
than that derived by \citet{1991ApJ...374..431S}. In this work, we
use \citet{pad07}'s radio LF to derive the $\gamma$-ray LF of BL Lac
objects.

Although a tight correlation between $L_{\rm R,5G}$ and $L_\gamma$
is found for $\gamma$-ray BL Lac objects, we find that only a small
fraction of BL Lac objects were detected by EGRET, which may imply
that the detection of blazars in $\gamma$-ray energy band is not
only dependent on their radio emission, but also on their
$\gamma$-ray activity. The $\gamma$-ray emission is closely related
with radio emission for BL lac objects based on SSC models. More
distant or/and fainter BL Lac objects can be detected if the
sensitivity is improved. \citet{a09a} found that the redshift
distribution of Fermi BL Lac objects ($z\leq 1.2$) is similar to
that of the DXRBS BL Lac sample ($z\lesssim 1.0$)\citep{pad07}. This
may imply that the radio-gamma correlation found in this work has
not been affected by the redshift distribution of $\gamma$-ray BL
Lac objects. Thus, we induce a $\gamma$-ray duty cycle
$\delta_{\gamma}$ for BL Lac objects, which can be estimated from
the ratio of $\gamma$-ray BL Lac objects to the total (e.g.,
\cite{ste96, ver04}). There are 36 BL Lac objects were detected by
EGRET and Fermi, among which the lowest radio flux density at 5~GHz
is $\simeq$0.25 Jy. The number of BL Lac objects above this flux
density limit with the same sky coverage of the survey carried out
with EGRET/Fermi is $\sim$330 (e.g., see figure \ref{fig4} in
\cite{pad07}), which leads to $\delta_{\gamma}\simeq 0.11$.
\citet{ver04} estimate the $\gamma$-ray activity using three
different methods. Our result is consistent with the maximum
distribution of blazars active fraction computed in \citet{ver04}
(see Fig.14 of \citet{ver04}).

The $\gamma$-ray LF of BL Lac
objects can then be calculated with
\begin{equation}
\Phi_{\gamma}(L_{\gamma},z)=\delta_{\gamma} \Phi_{\rm R}(L_{\rm
R,5G},z)\frac{{\rm d}L_{\rm
R,5G}}{dL_{\gamma}}=\frac{\delta_{\gamma}L_{\rm
R,5G}}{0.995L_{\gamma}}\Phi_{\rm R}(L_{\rm R,5G},z), \label{GLF}
\end{equation}
where $\Phi_{\gamma}$ is the $\gamma$-ray LF, and the radio LF
$\Phi_{\rm R}$ is given by \citet{pad07}. We can use a simple GLF
model defined as
\begin{equation}
\Phi_{\gamma}(L_{\gamma},z)\propto L_{\gamma}^{\beta}, \label{GLF2}
\end{equation}
In figure \ref{fig2}, we compared our $\gamma$-ray LF
\textbf{($\beta=-2.13$)} derived from the radio LF and that directly
derived with a sample of $\gamma$-ray BL Lac objects detected by
Fermi \citep{a09a}. The slope of our $\gamma$-ray LF is well in
agreement with the value of $-2.17\pm0.05$ reported for Fermi BL Lac
objects. Our $\gamma$-ray LF is slightly less than that of
\citep{a09a}, but taking into account the observational error, we
can also say our $\gamma$-ray LF is roughly consistent with that of
Fermi.

\begin{figure}
  \begin{center}
    \FigureFile(80mm,80mm){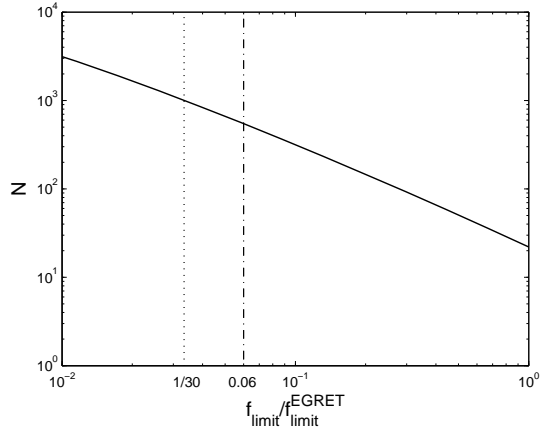}
  \end{center}
  \caption{The number count of $\gamma$-ray BL Lac objects as a
function of flux limit at 100~MeV.  The dotted line and dot-dashed
line represents the sensitivity of Fermi. We note $f_{\rm min}\simeq
5\times 10^{-8}\ ph~cm^{-2}~s^{-1}$ for EGRET (see, e.g.,
\cite{cao08}). The flux limit of Fermi at 100~MeV can be 30 times
lower than that of EGRET for a two year all-sky survey\citep{geh99}
and the sensitivity for one year will be $3.0\times10^{-9}ph\
cm^{-2}s^{-1}$\citep{atw09}.}\label{fig3}
\end{figure}

The number count of BL Lac objects as a function of $\gamma$-ray
flux above 100~MeV can be calculated with the derived $\gamma$-ray
LF,
\begin{equation}
N(\ge f_{\nu,\gamma}^{\rm min})=\int_0^{z_{\rm
m}}\frac{dV}{dz}dz\int\limits_{4\pi{d_{\rm L}^2}{f_{\nu,\gamma}^{\rm
min}}}\phi_{\gamma}(L_{\nu,\gamma},z)dL_{\nu,\gamma},  \label{n BL}
\end{equation}
where $z_{\rm m}=1$ is adopted in our calculations, because the
radio LF for BL Lac objects was derived with a sample of sources
with redshifts $z=0-1$ (see \cite{pad07}, for the details). This
means that the results derived here are only the lower limits. In
fact, almost all $\gamma$-ray BL Lac objects detected have redshifts
$z\lesssim 1$ (only 2 of 40 have redshifts slightly higher than
unity). We plot the number count of $\gamma$-ray BL Lac objects
derived from the LF calculated with equation (\ref{GLF}) in figure
\ref{fig3}. Obviously, the increasing number of sources can be
detected following a decreasing flux limit, such as $N\sim 500$ at
$f_{\rm limit}=0.06f_{\rm limit}^{EGRET}$ (the Fermi sensitivity for
one year) \citep{geh99}, and $N\sim 1000$ at $f_{\rm
limit}=1/30f_{\rm limit}^{EGRET}$ (the Fermi sensitivity for two
year) \citep{atw09}.



The EGRB contributed by all BL Lac objects can be calculated with
the $\gamma$-ray LF derived with equation (\ref{GLF}),
\begin{equation}
f_{\rm EGRB}=\frac{1}{4\pi}\int_0^{z_{\rm
m}}{dz}\int\limits_{4\pi{d_{\rm L}^2}{f_{\nu,\gamma}^{\rm
min}}}\frac{dN(L_{\nu,\gamma},z)}{dzdL_{\nu,\gamma}}\frac{L_{\nu,\gamma}(1+z)^{2-{\Gamma}}}{4\pi{d_{\rm
L}^2}E_{\rm 100MeV}}dL_{\nu,\gamma}, \label{EGRB}
\end{equation}
where the average photon spectral index $\Gamma=2.04$ for BL Lac
objects is adopted, which is slightly higher than $\Gamma=1.99\pm
0.22$ given by \citet{a09a}. This is due to that our sample includes
9 EGRET sources, of which the photon spectral indexes are all
greater than 2.0 except 1011$+$496. The contribution of the BL Lac
objects to the EGRB as a function of $\gamma$-ray flux limit $f_{\rm
limit}$ is plotted in figure \ref{fig4}.


\section{Discussion}


It was suggested that the $\gamma$-ray radiative mechanisms are
different for quasars and BL Lac objects, i.e., the EC mechanism may
be responsible for quasars, while the SSC is for BL Lac objects
(e.g., \cite{don95}). A linear relation between radio and
$\gamma$-ray emission is expected for AGNs, if the SSC mechanism is
responsible for $\gamma$-ray emission (see, e.g., equation (28) of
\cite{1997ApJS..109..103D}). The correlation between radio and
$\gamma$-ray emission for BL Lac objects found in this work is very
close to a linear one as described in equation(1), which seems to
support the SSC mechanism for $\gamma$-ray emission from BL Lac
objects.

The variability of $\gamma$-ray emission in 10 sources of 42 Fermi
BL Lac objects have been detected. The faction is much less than
that of FSRQs (45/57)\citep{a09a}, which implies that the
variability of $\gamma$-ray emission from BL Lac objects could be
weaker than that from FSRQs. The detection rate in $\gamma$-ray band
may be dominantly related to the duty cycle of $\gamma$-ray BL Lac
objects. \citet{ver04} investigated the $\gamma$-ray activity of
EGRET blazars and estimated their $\gamma$-ray duty cycle. They
found that about 48 percent of the sources fall into
$\delta_\gamma=0-5\%$, and 74 percent sources are in
$\delta_\gamma\le 10\%$.  The duty cycle $\delta_\gamma\simeq 0.11$
derived in this work is roughly consistent with the results in
\citet{ver04} (while their $\gamma$-ray duty-cycle was computed
 without a clear distinction between BL Lacs and FSRQs.).

\begin{figure}
  \begin{center}
    \FigureFile(80mm,80mm){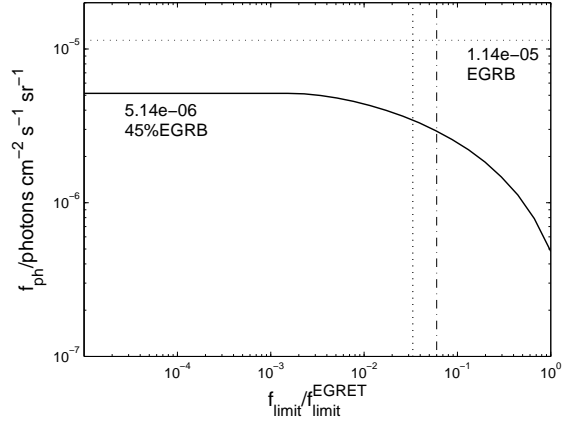}
  \end{center}
  \caption{The contribution of the BL Lac objects with $\gamma$-ray
flux $f\ge f_{\rm limit}$. The dashed line indicates the measured
EGRB with photon energy above 100~MeV. We use the value of the EGRB
obtained by \citet{str04}, $(1.14\pm 0.12)\times 10^{-5}$photons
cm$^{-2}$ s$^{-1}$ sr$^{-1}$. The symbols of dotted line and
dot-dashed line as in Figure \ref{fig3}.}\label{fig4}
\end{figure}

In this work, the $\gamma$-ray LF of BL Lac objects is derived from
the radio LF by using the empirical correlation of $\gamma$-ray
emission with radio emission, which can almost reproduce the
$\gamma$-ray LF calculated directly with the $\gamma$-ray BL Lac
objects detected by Fermi/EGRET (see figure \ref{fig2}). Based on
the derived $\gamma$-ray LF for BL Lac objects, we can calculate the
number count of $\gamma$-ray BL Lac objects as a function of flux
limit in $\gamma$-ray band. We find that about 1000 BL Lac objects
will be detected by Fermi, if its sensitivity is 30 times higher
than that of EGRET at 100~MeV \citep{geh99}. \citet{cao08}'s
estimate show that about 1200 quasars can be detected by Fermi based
on the EC mechanism for their $\gamma$-ray emission. The recently
released bright AGN sample detected by Fermi together with the
sources already detected by EGRET with high-confidence leads to a
sample of 135 $\gamma$-ray blazars (77 quasars, 51 BL Lac objects
and 7 blazars with uncertain classification) . The ratio of quasars
to BL Lac objects for the present sample (LBAS+EGRET) is about 1.5,
which is slightly higher than 1200/1000,
which implies that more BL Lac objects are expected to be detected
by the Fermi in the future.



Integrating the derived $\gamma$-ray LF, we calculate the
contribution of all BL Lac objects to the EGRB, which accounts for
$\sim$45\% of the EGRB (see figure \ref{fig4}). \citet{cao08}'s
calculation show that all radio quasars and FR II radio galaxies
contribute about $\sim$32\% of the EGRB. This means that the
contribution of BL Lac objects is similar to that of quasars, which
is roughly consistent with the predicted number ratio of quasars to
BL Lac objects to be detected by Fermi. The contribution of all
blazars (quasars+BL Lac objects) can account for $\sim$77\% of the
EGRB, which implies there still some space ($\lesssim 23\%$) is left
for other sources to the EGRB. Considering that our calculations are
limited to $z=0\sim1$ for BL Lac objects, the contribution of the
sources other than blazars to the EGRB should be lower than the
value derived in this work. We also noticed the detection of
$\gamma$-ray emission from relativistic jets in the narrow-line
Seyfert-1 galaxies (PMN 0948+0022)\citep{a09b}, which means that the
mechanism of $\gamma$-ray radiation from narrow-line Seyfert-1
galaxies is similar to that for blazars. Considering that a fraction
of narrow-line Seyfert-1 galaxies are radio-loud (e.g.
\cite{2006AJ....132..531K,2008ApJ...685..801Y}) and some of them
show evidence for relativistic jets (e.g.
\cite{2010arXiv1004.3058G}), the contribution of radio-loud
narrow-line Seyfert-1 galaxies to the EGRB may be important,
however, the detailed calculation is beyond the scope of this work.


\bigskip

 We thank the referee for his/her helpful comments, and J.~M. Bai for helpful discussion. This
work is supported by the NSFC (grants 10773020, 10821302 and
10833002), the CAS (grant KJCX2-YWT03), the Science and Technology
Commission of Shanghai Municipality (10XD1405000), and the National
Basic Research Program of China (grant 2009CB824800).

\newpage


\begin{longtable}{lccrrrccrr}
\caption{Literature Data for $\gamma$-ray BL Lac objects}
\hline
 IAU name &  source name &   z &     $F_{5GHz}$ &    $\Delta F_{5GHz}$ &   ref &   $\Gamma$ &  $\Delta\Gamma$ & $F_{100MeV}$ &  $\Delta F_{100MeV}$ \\
  &     &    & (Jy) & (Jy) &    &
   &   & $10^{-8}ph$ $cm^{-2}s^{-1}$ & $10^{-8}ph$ $cm^{-2}s^{-1}$\\

 \hline
\endfirsthead
\hline
 IAU name &  source name &   z &     $F_{5GHz}$ &    $\Delta F_{5GHz}$ &   ref &   $\Gamma$ &  $\Delta\Gamma$ &    $F_{100MeV}$ &  $\Delta F_{100MeV}$ \\
  &     &    & (Jy) & (Jy) &    &
   &   & $10^{-8}ph$ $cm^{-2}s^{-1}$ & $10^{-8}ph$ $cm^{-2}s^{-1}$\\
 \hline
\endhead
\hline
\endfoot
\hline \multicolumn{10}{l}{ } \\
\hline
\endlastfoot
$0048-097$ &    0FGL J$0050.5-0928$ &   0.537 &     1.92 &  ... &   3 &     2.15 &  0.08 &  10.2 &  1.4    \\
$0109+224$ &    0FGL J$0112.1+2247$ &   0.265 &     0.2575 &    0.0382 &    4 &     2.10 &  0.09 &  7.4 &   1.2    \\
$0118-272$ &    0FGL J$0120.5-2703$ &   0.557 &     1.18 &  ... &   3 &     1.99 &  0.14 &  3.2 &   0.8    \\
$0219+428$ &    0FGL J$0222.6+4302$ &   0.444 &     0.953 &     0.165 &     4 &     1.97 &  0.04 &  25.9 &  1.6    \\
$0235+164$ &    0FGL J$0238.6+1636$ &   0.940 &     1.48 &  0.15 &  5 &     2.05 &  0.02 &  72.6 &  2.5    \\
$0301-243$ &    0FGL J$0303.7-2410$ &   0.260 &     0.39 &  ... &   3 &     2.01 &  0.13 &  3.9 &   0.9    \\
$0332-403$ &    0FGL J$0334.1-4006$ &   1.445 &     2.6 &   ... &   3 &     2.15 &  0.12 &  5.3 &   1.1    \\
$0426-380$ &    0FGL J$0428.7-3755$ &   1.112 &     1.14 &  ... &   3 &     2.14 &  0.05 &  24.5 &  1.8    \\
$0447-439$ &    0FGL J$0449.7-4348$ &   0.205 &     0.2365 &    0.015 &     6 &     2.01 &  0.06 &  12.0 &  1.3    \\
$0502+675$ &    0FGL J$0507.9+6739$ &   0.416 &     0.025 &     ... &   4 &     1.67 &  0.18 &  1.7 &   0.8    \\
$0521-365$ &    EGR J$0529-3608$ &  0.055 &     9.23 &  ... &   3 &     2.63 &  0.42 &  17.57 &     6.8    \\
$0537-441$ &    0FGL J$0538.8-4403$ &   0.892 &     3.8 &   ... &   3 &     2.19 &  0.04 &  37.6 &  2.2    \\
$0716+714$ &    0FGL J$0722.0+7120$ &   0.310 &     1.12 &  0.01 &  2 &     2.08 &  0.05 &  16.4 &  1.4    \\
$0735+178$ &    0FGL J$0738.2+1738$ &   0.424 &     3.14 &  0.12 &  5 &     2.10 &  0.14 &  4.6 &   1.1   \\
$0814+425$ &    0FGL J$0818.3+4222$ &   0.530 &     0.956 &     ... &   10 &    2.07 &  0.08 &  9.6 &   1.3    \\
$0829+046$ &    EGR J$0829+0510$ &  0.180 &     0.700 &     ... &   3 &     2.47 &  0.40 &  16.53 &     5.1    \\
$0847-120$ &    EGR J$0852-1224$ &  0.566 &     0.834 &     0.045 &     7 &     1.58 &  0.58 &  23.2 &  11.1   \\
$0851+202$ &    0FGL J$0855.4+2009$ &   0.306 &     2.30 &  0.12 &  5 &     2.31 &  0.11 &  9.0 &   1.5    \\
$0954+658$ &    EGR J$0956+6524$ &  0.368 &     1.46 &  ... &   1 &     2.08 &  0.24 &  12.65 &     3.0    \\
$1011+496$ &    0FGL J$1015.2+4927$ &   0.212 &     0.146 &     ... &   10 &    1.73 &  0.07 &  4.9 &   0.7    \\
$1011+496$ &    EGR J$1009+4831$ &  0.200 &     0.146 &     ... &   10 &    1.90 &  0.37 &  5.63 &  1.7    \\
$1053+494$ &    0FGL J$1053.7+4926$ &   0.140 &     0.0427 &    ... &   9 &     1.42 &  0.20 &  0.5 &   0.3    \\
$1055+567$ &    0FGL J$1058.9+5629$ &   0.143 &     0.146 &     ... &   10 &    2.11 &  0.14 &  3.9 &   1.0   \\
$1101+384$ &    0FGL J$1104.5+3811$ &   0.030 &     0.421 &     0.06 &  10 &    1.77 &  0.04 &  15.3 &  1.1    \\
$1215+303$ &    0FGL J$1218.0+3006$ &   0.130 &     0.262 &     ... &   10 &    1.89 &  0.06 &  9.7 &   1.1    \\
$1219+285$ &    0FGL J$1221.7+2814$ &   0.102 &     0.6755 &    0.17 &  5 &     1.93 &  0.07 &  8.3 &   1.1    \\
$1334-127$ &    EGR J$1337-1310$ &  0.539 &     2.18 &  ... &   3 &     2.62 &  0.42 &  18.7 &  6.5    \\
$1514-241$ &    0FGL J$1517.9-2423$ &   0.048 &     1.907 &     0.17 &  5 &     1.94 &  0.14 &  4.1 &   1.2    \\
$1553+113$ &    0FGL J$1555.8+1110$ &   0.360 &     0.638 &     0.096 &     4 &     1.70 &  0.06 &  8.0 &   1.0    \\
$1604+159$ &    EGR J$1607+1533$ &  0.357 &     0.281 &     ... &   10 &    2.06 &  0.41 &  39.3 &  12.3   \\
$1652+398$ &    0FGL J$1653.9+3946$ &   0.033 &     1.1945 &    0.05 &  5 &     1.70 &  0.09 &  3.1 &   0.6    \\
$1717+178$ &    0FGL J$1719.3+1746$ &   0.137 &     0.94 &  ... &   3 &     1.84 &  0.07 &  6.9 &   0.9    \\
$1730-130$ &    EGR J$1734-1315$ &  0.902 &     4.10 &  ... &   3 &     2.23 &  0.10 &  32.54 &     9.3    \\
$1749+096$ &    0FGL J$1751.5+0935$ &   0.322 &     1.495 &     0.21 &  5 &     2.27 &  0.07 &  18.4 &  2.1    \\
$1803+784$ &    0FGL J$1802.2+7827$ &   0.680 &     2.8 &   ... &   5 &     2.25 &  0.14 &  6.0 &   1.4    \\
$1959+650$ &    0FGL J$2000.2+6506$ &   0.047 &     0.23 &  ... &   8 &     1.86 &  0.11 &  4.2 &   1.0    \\
$2005-489$ &    0FGL J$2009.4-4850$ &   0.071 &     1.19 &  ... &   3 &     1.85 &  0.14 &  2.9 &   0.9    \\
$2032+107$ &    EGR J$2032+1226$ &  0.601 &     0.77 &  ... &   3 &     2.83 &  0.26 &  13.05 &     3.0    \\
$2155-304$ &    0FGL J$2158.8-3014$ &   0.116 &     0.31 &  ... &   3 &     1.85 &  0.04 &  18.1 &  1.2    \\
$2200+420$ &    0FGL J$2202.4+4217$ &   0.069 &     2.49 &  ... &   5 &     2.24 &  0.12 &  8.5 &   1.8    \\

\end{longtable}

\end{document}